\documentclass{ws-p9-75x6-50}

\begin{document}

\title{Gravitational Energy-Momentum in MAG}

\author{James M. Nester,  Chiang-Mei Chen and Yu-Heui Wu}

\address{Department of Physics, National Central University,\\
Chungli, Taiwan 320, R.O.C.\\E-mail: nester@phy.ncu.edu.tw,
cmchen@phy.ncu.edu.tw}


\maketitle\abstracts{
    Energy-momentum (and angular momentum) for the Metric-Affine Gravity
    theory is considered from a Hamiltonian perspective (linked with the
    Noether approach).  The important roles of the Hamiltonian boundary
    term and the many choices involved in its selection---which give rise
    to many different definitions---are emphasized.  For each choice one
    obtains specific boundary conditions along with a value for the
    quasilocal, and (with suitable asymptotic behavior) total (Bondi and
    ADM) energy-momentum and angular momentum.  Applications include the
    first law of black hole thermodynamics---which identifies a general
    expression for the entropy.  Prospects for a positive energy proof are
    considered and quasilocal values for some solutions are presented.
}

\section{Gravitational Energy-Momentum}

Energy-momentum is a fundamental conserved quantity which is
associated with the symmetry of space-time geometry.  In the modern
view space-time geometry is dynamic and this is the basis for our
gravity theories.  The primary source of gravity is the
energy-momentum density for matter and {\em all other} interaction
fields.  But these sources can exchange energy-momentum with the
gravitational field locally, which leads to the expectation that
gravity should also have its own local energy-momentum density.

While total energy-momentum is well defined (for gravitating systems
with suitable asymptotics) standard techniques for identifying a local
``gravitational energy-momentum density'' gave only various,
noncovariant, reference frame dependent {\em
pseudotensors},\cite{CNC99} which cannot give a well defined
localization.  This can be understood in terms of the {\em equivalence
principle}, which implies that the gravitational field cannot be
detected at a point.

It is now believed that the proper idea is {\em quasi-local} (i.e.,
associated with a closed 2-surface) energy-momentum.  The many
proposals and approaches have been referred to
elsewhere.\cite{BY93,CN99} Amoung the various criteria for a good
quasilocal energy-momentum expression that have been advocated we
typically{\,}\cite{CY88} find good limits:  including ADM (spatial
infinity), Bondi (null infinity), weak field and flat spacetime.
However it has been observed that there are an {\em infinite number} of
expressions satisfying such requirements.\cite{Ber92} Hence additional
{\em principles} and {\em criteria} are very much needed.

\section
 {Hamiltonian approach}

One approach is to regard
energy as
the value of the Hamiltonian.
The gravitational Hamiltonian
(for a finite region $\Sigma$),
\begin{equation}
H(N)=\int_{\Sigma} N^\mu {\cal H}_\mu +
 \oint_{S=\partial \Sigma}{\cal B}(N),
\label{H(N)}
\end{equation}
depends on a displacement vector field $N$ and includes a spatial
hypersurface and a spatial 2-boundary term.  It turns out that the
boundary term plays a very important role, giving both the quasilocal
values and the boundary conditions.

For our purpose differential form notation has several advantages,
which are
associated with the (generalized
Stokes) boundary theorem, spacetime projection via pullback and the
interior product, and a neat representation of geometric objects.
Consider a {\em first order} Lagrangian for a k-form field
\begin{equation}
{\cal L}=d\varphi\wedge p-\Lambda.
\label{L1phi}
\end{equation}
The general variational formula
\begin{equation}
\delta {\cal L}=d(\delta\varphi\wedge p) +
 \delta\varphi\wedge {\delta {\cal L}\over \delta\varphi}+
 {\delta {\cal L}\over \delta p}\wedge \delta p
\end{equation}
implicitly defines the pairs of first order field equations.
Local diffeomorphism invariance requires this relation to be identically
satisfied for
$\delta=\pounds_N$, the Lie derivative.
Consequently, since
$\pounds_N\equiv i_Nd+di_N$ on the components of forms,
\begin{equation}
\pounds_N{\cal L}=d i_N {\cal L} \equiv d(\pounds_N\varphi\wedge p) +
 \pounds_N\varphi\wedge {\delta {\cal L}\over \delta\varphi}+
 {\delta {\cal L}\over \delta p}\wedge \pounds_N p.
\end{equation}
Hence the {\em Hamiltonian} 3-form,
\begin{equation}
{\cal H}(N):=\pounds_N\varphi\wedge p-i_N{\cal L},
\label{ham}
\end{equation}
satisfies the differential identity
\begin{equation}
d{\cal H}(N)
\equiv{\hbox{(terms proportional to field equations)}}.
\label{difident}
\end{equation}
Rearranging (\ref{ham})
using (\ref{L1phi}) gives an expression of the form
\begin{equation}
{\cal H}(N)=N^\mu{\cal H}_\mu+d{\cal B}(N).
\label{HdB}
\end{equation}
Upon substitution of
$d{\cal H}(N)=d(N^\mu{\cal H}_\mu)=dN^\mu\wedge{\cal H}_\mu+N^\mu
d{\cal H}_\mu$
into the differential identity (\ref{difident}),
 the coefficient of
$dN^\mu$ gives an algebraic identity
\begin{equation}
{\cal H}_\mu \equiv{\hbox{(terms proportional to field equations)}}.
\end{equation}
Hence ``on shell''
 (i.e., when the field
equations are satisfied)
 the Hamiltonian 3-form ${\cal H}(N)$ is a ``conserved
current''.
The ``conserved'' value of the Hamiltonian (the integral of
(\ref{HdB}), having the aforementioned form  (\ref{H(N)}))---
since
${\cal H}_\mu$
vanishes ``on shell''---
depends only on the
spatial 2-boundary term, which thus
 determines the quasilocal energy-momentum and angular
momentum.

However, as with other Noether currents, ${\cal H}(N)$ is not unique.
We can add to it a total differential (without changing the
Hamiltonian equations of motion or the conservation property).  This
amounts to modifying ${\cal B}(N)$, allowing one to ``improve'' the
quasilocal expression.  Indeed in many cases (including General
Relativity{\,}\cite{RT74}) it is {\em necessary} to adjust $\cal B$.
Fortunately, ${\cal B}$ is not arbitrary.  A further {\em principle}
of the formalism controls its form:  one should {\em choose} the
Hamiltonian boundary term ${\cal B}$ so that the boundary term in
$\delta H$ {\em vanishes}, when the desired fields are held fixed
(``controlled'') on $S$ (as discussed elsewhere{\,}\cite{RT74,CN01} in
detail, technically this is necessary in order for the Hamiltonian to
be differentiable).  There is thus a nice division:  the Hamiltonian
density ${\cal H}_\mu$ determines the evolution and constraint
equations, the boundary term ${\cal B}$ determines the {\em boundary
conditions} and the quasilocal energy-momentum.

Along with this Hamiltonian variation boundary principle we have
advocated an additional {\em criterion}, namely
{\em covariance}. For {\em each} dynamical field we found,\cite{CNT95,CN99}
using {\em symplectic} techniques,\cite{KT79}
 that
there are {\em only two} {\em covariant choices} for $\cal B$:
\begin{eqnarray}
{\cal B}_\varphi(N) &=&
 i_N\varphi \wedge \Delta p -(-1)^k \Delta \varphi \wedge
 i_N {\bar p},\\
{\cal B}_p(N) &=&
i_N {\bar \varphi} \wedge \Delta p
-(-1)^k \Delta \varphi \wedge i_N p,
\end{eqnarray}
here $\Delta \varphi:=\varphi-\bar\varphi$, and
$\Delta p:=p-\bar p$
  where
$\bar \varphi$, $\bar p$ represent reference values.
The associated Hamiltonian variations have the form
\begin{eqnarray}
\delta{\cal H}_\varphi(N)&=&
\hbox{field eqn terms} + d i_N (\delta \varphi \wedge \Delta p),
\\
\delta{\cal H}_p(N)&=&
\hbox{field eqn terms} - d i_N (\Delta \varphi \wedge \delta p),
\end{eqnarray}
revealing a boundary symplectic structure,
which yields the
 associated
boundary conditions: respectively
{\em Dirichlet} or {\em Neumann} ``{control mode}''.
Only for ${\cal B}_\varphi$ and ${\cal B}_p$ are the
Hamiltonian variation boundary terms
projections of 4-covariant expressions.
Note that, just as in thermodynamics
(with
 enthalpy, Gibbs, Helmholtz, \dots),
there are various kinds of energy
corresponding to different boundary conditions.

Specifying the quasilocal boundary term ${\cal B}(N)$ involves choices
including \begin{itemize}

\item the {\em representation}, (i.e., the {\em dynamic variables})
e.g., the metric, orthonormal frame, connection, spinors.

\item the {\em control mode}: the boundary conditions, essentially
Dirichlet or Neumann.

\item the {\em reference configuration}: e.g.,
Minkowski, de Sitter, Friedmann-Robertson-Walker cosmology, Schwarzschild.
 The meaning is that all quasilocal quantities {\em vanish}
when the field has the reference values,
so it determines the zero of energy etc.

\item the {\em displacement vector field} $N$:
Which timelike displacement gives the energy?
Which spatial displacement gives the momentum?
Which rotational displacement gives the angular momentum?

\end{itemize}

\section{Metric Affine Gravity}

We now apply these ideas to the Metric Affine Gravity Theory
(MAG).\cite{HMMN94,TW95,Gro97} The {\em geometric potentials} are the
 {\em metric coefficients}
$g_{\mu\nu}$, the  {\em coframe} 1-form
$\vartheta^\alpha$,  and the {\em connection} 1-form
$\Gamma^\alpha{}_\beta$.
The associated
{\em field strengths} are
\begin{eqnarray}
Dg_{\mu\nu}&:=&dg_{\mu\nu}-\Gamma^{\gamma}{}_{\mu}g_{\gamma\nu}
                        -\Gamma^{\gamma}{}_{\nu}g_{\mu\gamma},
\\
T^\alpha &:=& D \vartheta^\alpha := d \vartheta^\alpha +
\Gamma^\alpha{}_\beta \wedge \vartheta^\beta,
\\
R^\alpha{}_\beta &:=& d \Gamma^\alpha{}_\beta
+ \Gamma^\alpha{}_\gamma \wedge \Gamma^\gamma{}_\beta.
\end{eqnarray}
the
{\em non-metricity} 1-form,
the {\em torsion} 2-form, and the
{\em curvature} 2-form, respectively.

Independent variation with respect to the potentials
$(g,\vartheta,\Gamma)$ and conjugate momenta $(\pi,\tau,\rho)$
of the
``first order'' (source free) MAG Lagrangian 4-form:
\begin{equation}
{\cal L}
 :=Dg_{\mu\nu} \!\wedge\! \pi^{\mu\nu}
  + T^\alpha \!\wedge\! \tau_\alpha
  + R^\alpha{}_\beta \!\wedge\! \rho_\alpha{}^\beta
  - \Lambda( g, \vartheta; \pi, \tau, \rho),
\end{equation}
yields
\begin{eqnarray}
\delta {\cal L} &=&
  d ( \delta g_{\mu\nu} \pi^{\mu\nu}
     + \delta \vartheta^\alpha \wedge \tau_\alpha
     + \delta \Gamma^\alpha{}_\beta \wedge \rho_\alpha{}^\beta) \nonumber\\
&+& \delta g_{\mu\nu} \, {\delta {\cal L} \over \delta g_{\mu\nu}}
 + \delta \vartheta^\alpha \wedge
   {\delta {\cal L} \over \delta \vartheta^\alpha}
 + \delta \Gamma^\alpha{}_\beta \wedge
   {\delta {\cal L} \over \delta \Gamma^\alpha{}_\beta}  \nonumber\\
&
+&{\delta {\cal L} \over \delta \pi^{\mu\nu}} \wedge \delta \pi^{\mu\nu}
+{\delta {\cal L} \over \delta \tau_\alpha} \wedge \delta \tau_\alpha
+{\delta {\cal L} \over \delta \rho_\alpha{}^\beta} \wedge
   \delta \rho_\alpha{}^\beta,
\end{eqnarray}
which implicitly defines the {\em first order\/} equations
(the detailed form is not needed here).

We decompose the Lagrangian according to
\begin{eqnarray}
{\cal L} &\equiv& dt \wedge i_N {\cal L} \nonumber\\
&=& dt \wedge ( \pounds_N g_{\mu\nu} \pi^{\mu\nu}
  + \pounds_N \vartheta^\alpha \wedge \tau_\alpha
  + \pounds_N \Gamma^\alpha{}_\beta \wedge \rho_\alpha{}^\beta
  - {\cal H}(N) ).
\end{eqnarray}
to find the covariant {\em Hamiltonian} 3-form. Explicitly, it has the
standard
form (\ref{HdB})  where
 \begin{eqnarray}
 N^\mu{\cal H}_\mu &:=& i_N \Lambda
+ Dg_{\mu\nu} \wedge i_N
\pi^{\mu\nu}   - T^\alpha \wedge i_N \tau_\alpha
  - R^\alpha{}_\beta \wedge i_N \rho_\alpha{}^\beta
\nonumber  \\
 & -& i_N \vartheta^\alpha \wedge D \tau_\alpha
- i_N \Gamma^\alpha{}_\beta ( D \rho_\alpha{}^\beta
  - g_{\alpha\nu} \pi^{\beta\nu} - g_{\mu\alpha} \pi^{\mu\beta}
  + \vartheta^\beta \wedge \tau_\alpha ), \\
{\cal B}(N) &:=& i_N \vartheta^\alpha \tau_\alpha
  + i_N \Gamma^\alpha{}_\beta \, \rho_\alpha{}^\beta.
\label{MAGB1}
\end{eqnarray}
 We then replace the Hamiltonian boundary term (\ref{MAGB1}) by
choosing
one of the covariant  quasilocal boundary expressions for the MAG
\begin{eqnarray}
{\cal B}(N) =
\left\{ \matrix{  -\Delta g_{\mu\nu} \,
i_N{\overline{\pi}}{}^{\mu\nu}\cr
 -\Delta g_{\mu\nu} \, i_N \pi^{\mu\nu} \cr} \right\}
&+&
\left\{ \matrix{ \ \
 i_N \vartheta^\alpha \Delta \tau_\alpha &  \
   + \quad \Delta \vartheta^\alpha \wedge i_N
   {\overline{ \tau}}_\alpha \cr
 \ \ i_N {\overline{ \vartheta}}{}^\alpha
   \Delta \tau_\alpha & \
   + \quad \Delta \vartheta^\alpha \wedge i_N \tau_\alpha\cr} \right\}
\nonumber \\
&+&
\left\{ \matrix{
 {\tilde D}_\beta N^\alpha \, \Delta \rho_\alpha{}^\beta &
   + \quad \Delta \Gamma^\alpha{}_\beta \wedge i_N
   {\bar \rho}{}_\alpha{}^\beta \cr
  \overline{{\tilde D}_\beta N^\alpha}
  \, \Delta \rho_\alpha{}^\beta  &
   + \quad \Delta \Gamma^\alpha{}_\beta
   \wedge i_N \rho_\alpha{}^\beta \cr} \right\},
\end{eqnarray}
where the upper (lower) line in each bracket is to be selected if the
field (momentum) is controlled.  Again there are several kinds of
energy, each corresponds to the work done in a different (ideal)
physical process.

A technical point here is that
we replaced the $i_N\Gamma$ terms using the identity
\begin{equation}
\pounds_N\vartheta^\alpha\equiv
 D N^\alpha +i_N T^\alpha
-i_N\Gamma^\alpha{}_\beta \vartheta^\beta\equiv
{\tilde D} N^\alpha
-i_N\Gamma^\alpha{}_\beta \vartheta^\beta,
\end{equation}
and then dropped the non-covariant, frame gauge dependent
$\pounds_N\vartheta\Delta\rho$ terms, to obtain fully covariant
expressions.  These covariant end results follow directly from a
different treatment of the connection.\cite{CN01}

With standard flat asymptotics:
\begin{eqnarray}
N^\alpha &\sim& (\hbox{constant} + O(1/r) )^+ +
(\epsilon^\alpha{}_\beta x^\beta + O(1) )^-, \\
\{ \Delta g, \Delta \vartheta, \Delta \rho\} &\sim&
O^+(1/r) + O^-(1/r^2),\\
\{\Delta \pi, \Delta \tau, \Delta \Gamma\} &\sim&
O^-(1/r^2) + O^+(1/r^3),
\end{eqnarray}
we obtain asymptotically (at spatial infinity) finite values for the
quasilocal quantities and an automatically vanishing boundary term in
$\delta\int{\cal H}(N)$.

These quasilocal expressions have good correspondence limits to
special cases of the MAG including
GR,\cite{BY93,RT74,BO87,Chr88,Sor88,Kat85,KBLB9x} the Poincar{\'e}
Gauge Theory,\cite{BV88,Kaw88,Hec95} and the teleparallel
theory.\cite{Mol61}  The latter has recently had a
revival,\cite{Mal9x,AGP00} largely because of new hopes for its
utility regarding energy-momentum localization.

\section{Applications}

In conclusion we here briefly consider several applications of our MAG
Hamiltonian boundary term quasilocal energy-momentum expressions.

\begin{itemize}

\item
 {\em Black hole thermodynamics:}
By choosing the boundary on the horizon and at infinity we get the
first law and a generalized expression for the entropy:{\,}\cite{CN99}
\begin{equation}
T\delta S=\oint_H\kappa\epsilon^{\alpha\beta}\delta \rho_{\alpha\beta},
\end{equation}
where $\kappa$ is the surface gravity and $\epsilon^{\alpha\beta}$ is
the binormal to the horizon.

\item {\em A positive energy proof?}
The formalism gives the necessary expressions, but
one must consider each distinct parameter choice separately.
The prospects are very poor in general, but not bad
for a few special cases with limited $R^2$ terms,
  e.g., just $R^\alpha{}_\alpha$ or the scalar or pseudoscalar curvature
squared.

\item {\em Positive total energy test:} This test,\cite{PET} based on
the fundamental requirement that gravity should be purely
attractive, is
expected to give severe constraints on the parameters---in
principle---however it requires a lot of effort to get a result.

\item {\em Quasilocal quantities for exact solutions:}{\,}\cite{MAGII}

We calculated the quasilocal energy for some exact MAG solutions.
For the first solution found by Tresguerres{\,}\cite{Tre95} and the
first solution we found{\,}\cite{HCN97},
 the frame (with vanishing cosmological constant for simplicity)
is
\begin{equation}
\vartheta^0=fdt, \quad \vartheta^r=f^{-1}dr,\quad \vartheta^\theta=rd\theta,
\quad\vartheta^\varphi=r\sin\theta d \varphi,
\end{equation}
where $f^2=1-2m/r+b_{10}N_0^2/(2 \kappa a_0 r^2)$.  Details of the
necessary
parameter restrictions and expressions for the torsion and nonmetricity are
given in the cited works.  Using a Minkowski
reference
geometry and analytic matching (the simplest but probably not the most physical
choice)
for the quasilocal energy we found, for Dirichlet and Neumann boundary
conditions respectively,
\begin{eqnarray}
E&=&a_0r(1-f^{-1})+bf'(f^2-1)(1+n_0^2f^{-2}),   \\
E&=&a_0r(f-1)+2bf'(1-f)(f+n_0^2f^{-2}),
\end{eqnarray}
for our solution and
\begin{eqnarray}
E&=&a_0r(1-f^{-1})+bf'(f^2-1)(1+n_0^2f^{-2})+bn_0^2f^{-2}(1-f)/r,\\
E&=&a_0r(f-1)+2bf'(1-f)(f+n_0^2f^{-2})+bn_0^2(f^{-1}-1)/r,
\end{eqnarray}
for the Tresguerres solution. For the interesting special
case{\,}\cite{MMS98} which has $f=1-m/r$, we found
 \begin{equation}
E=a_0r(1-f^{-1}) \qquad \hbox{and}\qquad E=a_0r(f-1),
\end{equation}
for Dirichlet and Neumann boundary conditions respectively.  These few
examples are representative of our findings for other solutions.  All
of our results have the expected asymptotic limit:  $-a_0m$; however
much more work will be needed to appreciate the physical significance
of the detailed shape of these quasilocal energy distributions.

\end{itemize}

\section*{Acknowledgments}
This work was supported by the National Science Council of the R.O.C.
under grant number NSC 89-2112-M-008-020.

\eject

\end{document}